\documentclass[onecolumn,authoryear]{els-mrw} 

\usepackage{amsmath,amssymb,amsfonts,amsthm,makeidx,graphicx}
\usepackage{txfonts}
\usepackage{helvet}
\usepackage{hyperref}

\begin{document}

\chapter{White dwarf systems: exoplanets and debris disks}\label{chap1}

\author[1]{Uri Malamud}%

\address[1]{\orgname{Technion Israel Institute of Technology}, \orgdiv{Department of Physics}, \orgaddress{Technion City, Haifa 3200003, Israel}}
%\address[2]{\orgname{Name of Institute}, \orgdiv{Division or Department}, \orgaddress{Address of Institute}}

\articletag{Chapter initial draft}

\maketitle

%\begin{glossary}[Glossary]
%\term{Europe} the model is a coherent view of capital markets data that allows users to interact with the content in a consistent manner.

%\term{Primates} regardless of the source. Essentially, of sources. Properly deployed.

%\end{glossary}

\begin{glossary}[Nomenclature]
\begin{tabular}{@{}lp{34pc}@{}}
WD &White Dwarf\\
IR &Infrared\\
PR &Poynting-Robertson\\
\end{tabular}
\end{glossary}

\begin{BoxTypeA}[chap1:box1]{\large{Key Points}}
	\begin{itemize}
		\item[-]{Although a large fraction of white dwarfs are polluted, only a few \% contain other evidence for harboring planetary systems.}
		\item[-]{There are five discovered polluted white dwarf exo-planets, found at widely different separations.}
		\item[-]{The orbital evolution of exo-planets and smaller planetesimals depends on each system's planetary and stellar architecture.}
		\item[-]{When planetesimals are injected towards the vicinity of white dwarfs, they are disrupted by gravitational tides, forming disks.}
		\item[-]{Tidally formed debris disks are initially highly eccentric.}
		\item[-]{Observational techniques to characterize disks include infrared and X-ray measurements, gas detection and transits.}
		\item[-]{Observations point to highly diverse configurations of debris disks, from highly eccentric to compact, and undergoing changes.}
		\item[-]{Three categories of models are so far suggested in order to explain the broad and complex variety of observations.}
	\end{itemize}	
\end{BoxTypeA}

\begin{abstract}[Abstract]
Although there is abundant and diverse observational evidence in support of white dwarf stars hosting planets or debris disks which form in the catastrophic destruction of various planetary bodies, the key processes that explain these observations are still being intensely investigated. The study of white dwarf planetary systems offers a unique perspective on exo-solar composition, that cannot be obtained by any other means. This chapter describes the various observational techniques that are used in order to find and characterize exo-planets and debris disks around white dwarfs. In turn, it discusses how to theoretically interpret these observations by surveying an array of various research tools and models currently employed in this field.
\end{abstract}

\section{Introduction}\label{S:Intro}
Several billion years into the future, our Sun will exhaust its fuel. It would expand, turning into a giant star, eventually shedding its outer layers to become a white dwarf (WD) -- a perpetually fading remnant of its former glory.

We will not be around to see this -- planet Earth is likely to be engulfed by the star's expansion. Other planets, moons, asteroids and comets in the outer Solar system will be baked by intense radiation. Their orbits will expand, giving rise to a rich dynamical chaos. In the aftermath of this calamity, some of these objects will survive, but many unfortunate ones would be injected into tidal crossing orbits of our Sun's ultra-dense successor, the WD. As they do, they will be violently and repeatedly ripped apart, breaking into their smallest constituent building blocks, while gradually forming disks of debris. 

While we cannot hope to glimpse our own future, nature has given us a unique opportunity to revel in this great cosmic spectacle, in which we obtain vast amounts of information from the regrettable demise of other, less fortunate exo-planetary systems. WDs are therefore key players in a great cosmic show. They are important because they play the equivalent role of astrophysical mass spectrometers - possessing the unmatched ability to probe the composition of exo-Solar planetary systems.

WDs are compact stars, the size of a terrestrial planet, and yet they contain a very large mass, comparable and up to that of our Sun. Hence, their surface gravities are several orders of magnitude higher than that of our Sun, which will quickly drag inwards any elements in their atmosphere, heavier than its primordial hydrogen or helium. Nevertheless, between 25-50\% of WD atmospheres are indeed polluted with heavier elements \citep{KoesterEtAl-2014}, which could only have come from accreting planetary remains. By measuring the relative abundances of the heavy elements detected in their atmospheres, one can infer the chemical composition of these pollutants \citep{HarrisonEtAl-2021}. There is currently no other technique which gives us this unique ability to probe the bulk composition of other exo-solar planetary systems, enabling a distinctive comparison to material collected in our own backyard, such as meteorites or space mission sample returns. The connections we find by such means have profound astrophysical implications.

Thus far, not a single WD has ever been observed to vary its atmospheric elemental abundances despite decades of observations, suggesting that the accretion is mediated by a disk. Yet, neither the architecture of such disks, nor the sequence of events leading to such acceretion - are currently understood. Contemporary research focuses on whether the pollutants are single or multiple bodies; in what circumstances are they perturbed to tidal crossing orbits; how do the disks form and evolve; why do disks often evade detection; and what drives accretion from the disk to the WD atmosphere.

Some of these questions will be addressed in this chapter, surveying the current state of knowledge in the field. The chapter begins by discussing known surviving exo-planets around WDs. These planets shed light on and provide context for understanding the evolution of planetary architectures, which on other occasions inject planetesimals towards close proximity of the WD, bringing about the onset of debris disks formation. Then, multiple observational clues are presented with which we can interpret these debris disks. Finally, the different theoretical models are surveyed. These models have all been proposed in an attempt to match the incredible wealth and diversity of observed data. explore

\section{White dwarf exo-planets}\label{S:Exoplanets}
The last couple of decades have brought about a revolution in astrophysics. New space telescopes and techniques have allowed us to identify thousands of confirmed exo-solar planets. Besides their significance in the search for life beyond Earth, exo-planets are important primarily due to their diversity, which helps us gain critical and fundamental lessons regarding planet formation and evolution. Nevertheless, out of thousands of known exo-planets, there are currently only five which are known to orbit WDs. This is perhaps not surprising, given how small and therefore dim WD stars are, which makes finding their exo-planets more challenging.

The five known planets are listed in Table \ref{Tab:Exo-planets}, sorted in descending order according to their orbital distance from the WD. The planet around WD J0914+1914 stands out in having its presence inferred, rather than directly observed. This comes from the fact that WDs shine brightest when they form. Newly formed WDs could have an effective photospheric temperature in excess of 100000 K, however since they no longer burn up fuel, they gradually cool, as their cooling rate decreases rapidly in time. WD J0914+1914 is a very young WD, and therefore it has a large effective temperature of nearly 30000 K, and is capable of evaporating the atmosphere of its closely orbiting planet. Its planet is thus inferred by the existence of the WD gaseous disk and its particular properties, after having ruled out other possibilities \citep{GansickeEtAl-2019}.

\begin{table}[t]
	\TBL{\caption{Known exo-planets around white dwarfs}\label{Tab:Exo-planets}}
	{\begin{tabular*}{\textwidth}{@{\extracolsep{\fill}}@{}lll@{}}
			\toprule
			\multicolumn{1}{@{}l}{\TCH{designation}} &
			\multicolumn{1}{c}{\TCH{approximate orbital separation [au]}} &
			\multicolumn{1}{l}{\TCH{discovery technique}}\\
			\colrule
			WD 0806-661 b & 2500 & imaging\\
			PSR B1620-26 (AB) b & 23 & pulsar timing\\
			MOA-2010-BLG-477L b & 2.8 & microlensing\\
			WD J0914+1914\footnotemark{a} & 0.07 & spectroscopy\\
			WD 1856+534 b & 0.02 & photometry (transit)\\
			\botrule
	\end{tabular*}}{%
	\begin{tablenotes}
		\footnotetext[a]{Here the designation refers to the WD, not its planet, since the detection is indirectly inferred from the WD surrounding gas}
	\end{tablenotes}
	}%
\end{table}

Note first the vast range in star-planet separations, spanning 5 orders of magnitude. Planet MOA-2010-BLG-477L b represents the dividing line between two diametrically opposed dynamical histories. At a distance of 2.8 au, planets on significantly closer orbits than MOA-2010-BLG-477L b could not possibly have survived the expansion of the host star before it became a WD. Therefore, they must have originally orbited much further out, and either migrated inwards due to interaction with the pre-WD gaseous surrounding (without getting fully engulfed), or were at a later stage scattered to close proximity of the star, experiencing orbital settling due to tides. Beyond the orbit of MOA-2010-BLG-477L b, second-generation planets could potentially form, as one possibility. At very large separations, such as in the case of WD 0806-661 b, other mechanisms may be invoked, such as scattering by other planets or gravitational capture. In conclusion, the aforementioned WD systems exemplify the rich realm of possibilities for perturbing and modifying planetary configurations. The following section discusses how related dynamical paths can trigger and shape the formation of debris disks.

\section{Delivering planetesimals to tidal crossing orbits}\label{S:Delivering}
All WDs may in principle be polluted directly, by having planetesimals plunge straight into their atmospheres. However, the small radii of WD stars makes this possibility unlikely. The more likely alternative is entering the WD's Roche radius, which is the distance where strong tidal forces exerted by the WD overwhelm the forces keeping planetesimals together (self-gravity for large or internal cohesive strength for small planetesimals). The approximate location of the Roche radius is around 1 R$\odot$, the Sun's radius, and equivalent to about 100 WD radii or 0.005 au ($\sim$1/4 of the separation of WD 1856+534 b). An orbit interior to the Roche radius is also referred to as a tidal-crossing orbit. The cross section for tidal disruptions is thus understandably much greater than that of direct WD impacts, and many studies show that disruption is the fate awaiting nearly all planetesimals. In what follows, a description of various dynamical channels to deliver planetesimals into the Roche radius is provided. As in the case of exo-planets, these dynamical modes are dictated by planetary architectures.

\begin{figure}[h!]
	\centering
	\includegraphics[width=.75\textwidth]{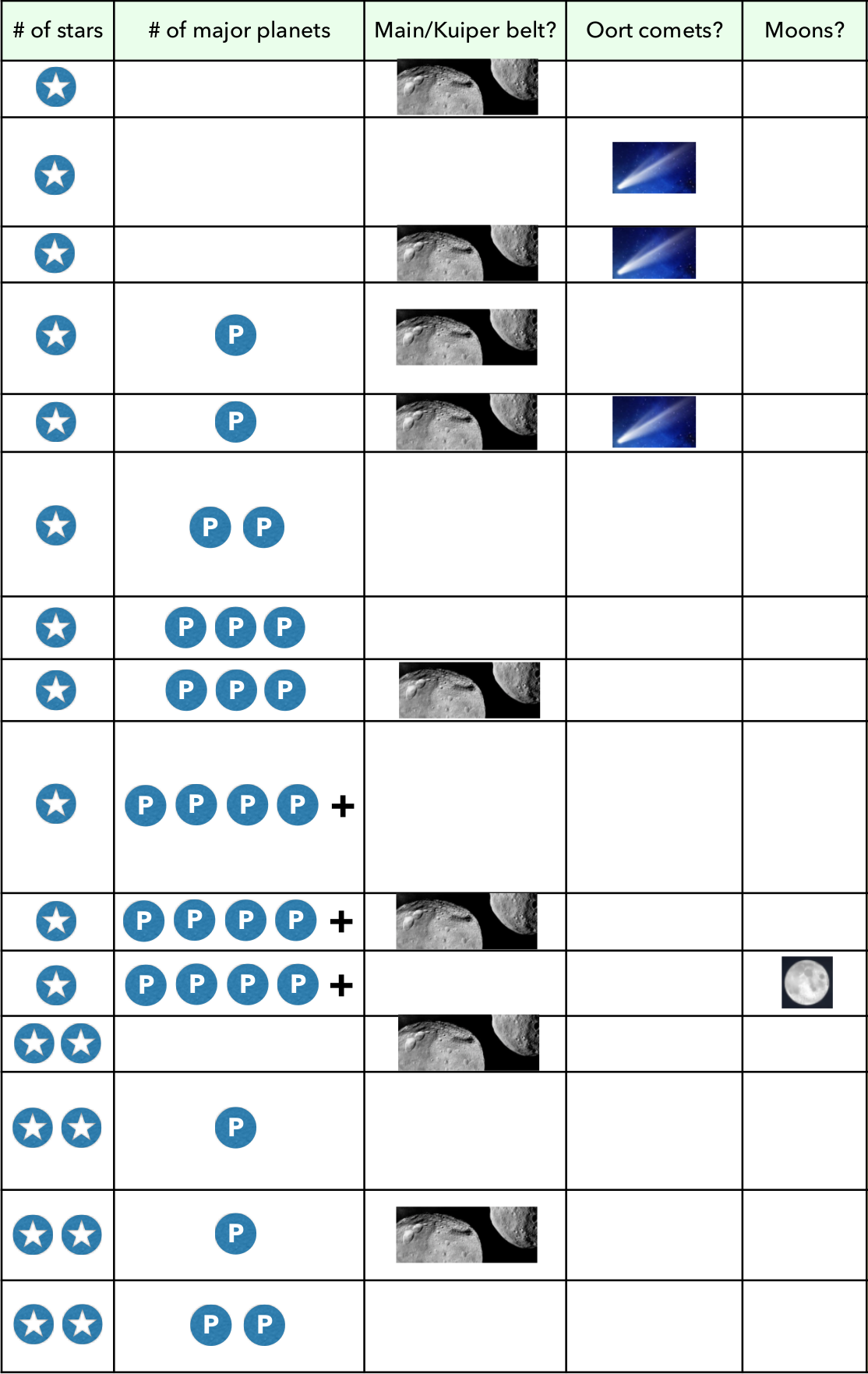}
	\caption{Planetary architectures leading to delivery of planetesimals and the formation of debris disks, detailing the number of stars and planets which were investigated by different studies, and the class of affected planetesimals (Credit: \cite{Veras-2021}, Figure 6).}
	\label{fig:Delivery}
\end{figure}

As discussed at great length in other chapters, typical polluters are inferred to be 'dry' bodies, hence asteroids-like in their composition. Moreover, the total mass of accreted material inside the atmospheres of WDs never exceeds that of small dwarf planets, and is typically similar to the mass of ordinary asteroids \citep{Veras-2016}. Therefore, analogues of solar system asteroids are the most likely progenitors of the majority of WD debris disks. Since asteroids are essentially massless in comparison to large planetary objects, it has been shown that gravitational interactions with giant planets, terrestrial planets and even large moons \citep{VerasRosengren-2023}, may easily perturb them into highly eccentric, tidal-crossing orbits. The eccentricity $e=(1-q/a)$ must be extremely high since any asteroid surviving engulfment requires an orbit with a semi-major axis $a$ larger than a few au, however the pericenter distance $q$ is necessarily within the Roche radius (<0.005 au), and therefore $e\approx1$.

While the initial trigger for pollution is likely to be the stellar mass-loss that precedes the formation of the WD, which widens planetary orbits and can destabilize planetary systems, the delivery of planetesimals was found to be ongoing and could last for many billions of years. Past numerical studies considered various planetary configurations, initially investigating the effect of a single planet, and then gradually moving towards an increasing number of planets. Asteroids are shown to pollute WDs across these various architectures.

Although there are far less polluted WDs that have accreted comet-like material, which is more rich in volatile ices and organics, a few cases are indeed known. Unlike asteroids, comets must originate from vast distances of hundereds and even thousands of au, at least if they are to retain their ices during the high-luminosity evolutionary phases of their host star, prior to becoming a WD \citep{MalamudPerets-2017}. Gravitational perturbations at large distances therefore include primarily stellar flybys and the potential influence of galactic tides.

Other possible polluters are larger bodies such as dwarf planets, moons, planets and giant planets. While they contain most of the mass in planetary systems, asteroids or comets greatly outnumber them by many orders of magnitude. Indeed, the observed mass accreted onto WD atmospheres is severely limited \citep{Veras-2016}, and in all cases lies below the mass of such large bodies, which may constitute as an empirical confirmation to their rarity. Nevertheless, theoretical studies show that their mutual gravitational interactions can in principle lead to scattering, and the eventual (extremely) rare accretion of a few of them. 

All categories of objects thus far mentioned -- asteroids, comets and larger bodies, can also be scattered due to the gravitational influence of a stellar companion to the WD. A breakdown of potential planetary and stellar configurations leading to the delivery of pollutants is comprehensively reviewed by \cite{Veras-2021}, and presented in Figure \ref{fig:Delivery}.

The following section discusses how the planetesimals that are delivered to tidal crossing orbits are torn apart and form an initial disk of debris.

\section{Tidal disruption initial sequence}\label{S:Disruption}
\begin{figure}[h!]
	\centering
	\includegraphics[width=.85\textwidth]{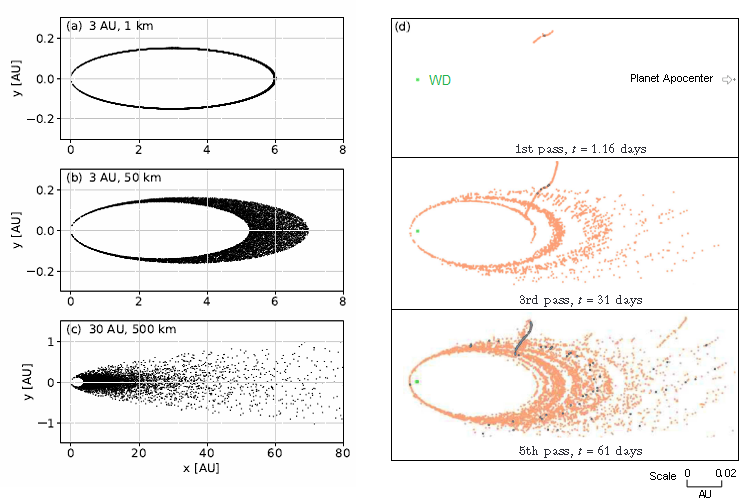}
	\caption{Debris disk formation and evolution following tidal disruptions: (a) small size (1 km) and semi-major axis (3 AU) results in a ring-like disk; (b) increasing size (50 km) results in some dispersion of the fragment orbital energies; (c) increasing size (500 km) and separation (30 AU) result in a bi-modal disk where half of the tidal fragments are unbound and the other half are tightly bound to the WD (Credit: \cite{BrouwersEtAl-2022}, Figure 4); (d) A full hydrodynamical evolution of the tidal disruption of an Earth-sized planet, leading to a bimodal disk. For numerical reasons the planet apocenter is unrealistically placed at 0.2 au (top right). Each flyby of planetary remains near the WD generates new tidal disruptions (top to bottom sequence). Light and dark pixel tones indicate silicate mantle and iron core composition, respectively (Credit: \cite{MalamudPerets-2020}, Figure 7).}
	\label{fig:Disruption}
\end{figure}

Now the unavoidable pull of the WD brings forth the final blow to the planetesimal's existence as a single entity. Upon reaching the Roche sphere, it is being differentially pulled and stretched from both sides. Gradual elongation soon turns to complete breakup. Where once before its constituent building blocks traveled jointly, now they each travel with the same velocity as before, while experiencing a slightly different gravitational tug by the WD. Their post-breakup spatial distribution is therefore translated into a spread in orbital energies. This process is what one refers to as a tidal disruption. Debris disks are formed by repetitive sequences of tidal disruptions, and their ultimate shapes are fundamentally determined by the size and origin of their progenitor planetesimals, which in turn determine the orbital energies of the debris \citep{MalamudPerets-2020}. When the progenitors have small sizes and semi-major axes, the debris remain fully bound to the WD and fill out a narrow ring along the original progenitor orbit. When they are either very large or originate from a large distance, a bimodal disk ensues where half of the debris are launched into interstellar space, and the other half become tightly bound to the WD. All other cases fall between these two extremes (Figure \ref{fig:Disruption}).

The incipient debris disks that emerge from tidal disruptions are only the harbinger of a much longer life cycle. From here on, various forces shape and govern the continued evolution of the disk, and the eventual accretion onto the WD atmosphere. Before discussing some of these processes, the following section first describes multiple and diverse observations, providing scientists with crucial clues as to the nature of these disks and the transformations they undergo.

\section{Observations of debris disks}\label{S:Observations}
There are numerous methods with which to observe, identify and characterize WD debris disks. In the following they are discussed one by one. Figure \ref{fig:Observations} then shows the first system to have been identified via each respective observation.
 
\subsection{Infrared excess}\label{SS:IRexcess}
\begin{figure}[h!]
	\centering
	\includegraphics[width=.95\textwidth]{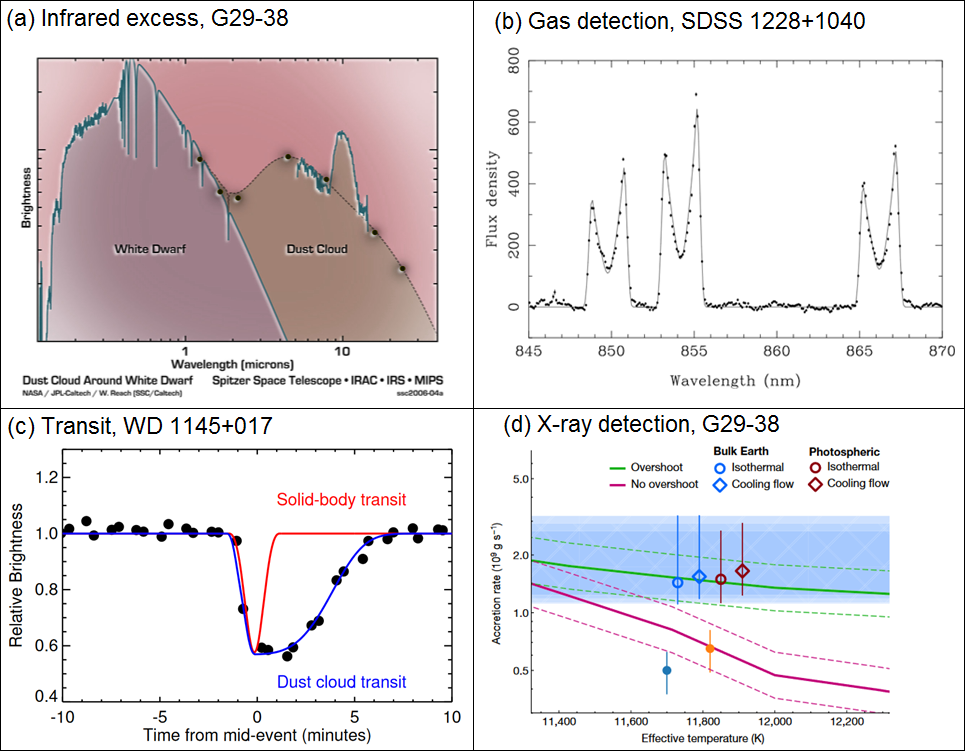}
	\caption{Listed are the seminal systems to be identified by each respective observational method. Credit: (a)  NASA/JPL-Caltech/M. Kuchner (GSFC); (b) \cite{GansickeEtAl-2006}; (c) \cite{VanderburgEtAl-2015}; (d) \cite{CunninghamEtAl-2022}.}
	\label{fig:Observations}
\end{figure}

While up to half of all WDs have atmospheres polluted with heavy elements, with thousands currently known, it might be surprising that only a small fraction between 1\%-3\% possess detectable circumstellar debris \citep{BonsorEtAl-2017,RebassaMansergasEtAl-2019}. Spectral observations of the first polluted WD with a debris disk, G29-38, confirmed the presence of silicate dust, based on characteristic silicate spectral features (see Figure \ref{fig:Observations}-a at 10 $\mu$m wavelength). The dust is registered through infrared (IR) excess, as the measured IR flux exceeds what is expected from the WD alone. Circumstellar dust is heated by stellar light and then re-emitted at longer IR wavelengths, with a characteristic black body temperature of around 1000 K \citep{Farihi-2016}, leading to the observed excess. There are now approximately 60 dusty systems known \cite{Veras-2021}, and the number is expected to rise dramatically in the next few years.

Since dusty disks are expected to form following tidal disruptions, warm dust is anticipated to fall within the Roche radius. In the classic technique for modelling IR excess \citep{Jura-2003}, one assumes that the dust is enclosed within a passive, opaque, geometrically flat disk that lies close to the WD, analogous to the rings of Saturn. This yields a typical declining disk temperature as a function of radial distance. In turn, it constrains the flux, and when compared with the observations, the inner and outer radii of the disk can be extracted, although there is some degeneracy between the disk inclination and its radial extent. Studies infer various disks to reside between $\sim$0.2-1.2 $R_\odot$.

Such models were favoured for at least a decade, placing the dust within the expected spatial range of the Roche radius. There are however at least three exceptional systems which cannot be fitted by the classic model of flat and opaque disks. These systems are GD 56, GD 362 and WD J0846+5703. For these especially bright systems, extended models were suggested in order to fit the IR excess, including an outer, optically thin dusty component \citep{JuraEtAl-2007}.

\subsection{Gas}\label{SS:Gas}
The detection of gas around polluted WDs is rarer than the detection of dust. With one exception (WD J0914+1914 discussed in Section \ref{S:Exoplanets} is a one-of-a-kind no dust system), gas is \emph{always} found as an additional component in dusty disks.

By the most recent count, there are some 26 known WD disks that contain a gaseous component. In 21 of the 26 aforementioned disks the presence of gas has been detected via line emission. In the other five disks absorption features are detected. The occurrence rate for dust disks with an observable gaseous component may therefore be up to about 1/3. The dust and gas therefore appear to be related, and the relatively small fraction of gas occurrences might either be fundamental or simply the result of most systems falling below the detection limit.

Some spatial constraints are possible by modelling systems with gas emission. Significant asymmetry in their line profiles indicate non-negligible eccentricity, ranging between $\sim$0.2-0.4 \citep{MelisEtAl-2010}. If modelled via a series of co-aligned elliptical orbits of identical eccentricity, the radial distribution of the gas may be constrained, placing it around 0.15-1.2 $R_\odot$ \citep{GansickeEtAl-2006,MelisEtAl-2010}. 
Absorption features thus far provide spatial constraints for only one system - in WD 1145+017 the gas is placed at $\sim$0.2-0.5 $R_\odot$\citep{FortinArchambaultEtAl-2020}.

In several systems, line morphology over long periods is commensurate with relativistic apsidal precession, due to the strong gravity of the WD, which periodically distorts the disk's orientation \citep{ManserEtAl-2016,FortinArchambaultEtAl-2020}. Constraints for the precession periods range from 4.6 yr for WD 1145+017 \citep{FortinArchambaultEtAl-2020} to 27 yr for SDSS J1228+1040 \citep{ManserEtAl-2016}. \cite{ManserEtAl-2016} find that precession periods of 1.54, 27.8 and 134 yr correspond to orbits with semi-major axes of 0.2, 0.64 and 1.2 $R_\odot$, respectively, suggesting that gas located in the inner parts of these disks might be involved.

Finally, one unique system (SDSS J1617+1620) was identified to host a transient gaseous disk, which following its first detection in 2008, monotonically decreased in strength over eight years until it dropped below detectability.

\subsection{Infrared variability}\label{SS:IRVariability}
Past observations in 2010 showed that the IR flux in WD J0959-0200 has decreased by 35\% in less than 300 days \citep{XuJura-2014}. Similarly, SDSS J1228+1040 decreased by 20\% between 2007 and 2014, to a level also seen in 2018, and G29-38 increased by 10\% between 2004 and 2007 \citep{XuEtAl-2018}. This kindled the idea that WDs with IR excess could be changing. More recent 3 yr photometric near-IR Earth-based survey of 34 WDs showed no significant variability \citep{RogersEtAl-2020}. Nonetheless, space-based observations at larger wavelength than the latter study are much more sensitive (to cooler dust), and it was recently shown (see \cite{SwanEtAl-2020} and references therein) that IR variability is unequivocally found across a large population of WD systems. The typical IR flux changes between several \% to tens of \%. The time scale of variability is months rather than hours or days. Interestingly, the highest variability correlates with WD systems which also feature emitting gas.

Two additional systems are especially important, as they provide some further temporal detail. The first system is WD 0145+234. This system brightened considerably in 2018, the IR flux tripling over the course of $\sim$1.5 years. This outburst in IR was followed-up by subsequent observations during late 2019, showing instead a steady linear decrease in the flux, however with small occasional spikes in brightness \citep{SwanEtAl-2021}. Recent observations show this system returned to its pre-outburst state. The second system is GD 56, featuring a similar declining flux, however one which lasted several years, rather than several weeks. An actual outburst might have preceded the declining flux however observations were not made at the time. Several years prior, the baseline IR flux of the system was measured to be similar to that at the end of the decline phase \citep{SwanEtAl-2021}.

\subsection{Transits}\label{SS:Transits}
Since WDs are physically small, it is expected that transiting material could block a significant fraction of their light when passing in front of the stellar disk, leading to much deeper transits than those of typical main sequence stars. Such transits were only observed relatively recently, and a total of 8 transiting WD systems are now known. The first system \citep{VanderburgEtAl-2015} exhibited transit periodicity of 4.5 hours, having an orbit in surprising close proximity to the Roche radius. The transits were deep, and at maximum could exceed 50\% of the WD's light. They were not taken to be the result of an eclipsing solid body, but rather predominantly of a dusty effluence, and were interpreted as a disintegrating asteroid. 

This followed by discovery of transits around ZTF J0139+5245, with similarly deep light curves, however the periodicity was around 107 days, commensurable with a semi-major axis of 0.355 AU. If the transiting material passes within the Roche radius, the eccentricity is constrained to be at least 0.97. The transit durations last several weeks. At periastron, a transiting grain would take $\sim$1 minute to transit, and around $\sim$1 hours if at apastron. The much longer transits observed, imply the presence of an extended stream of disrupted debris.

Five more transiting systems were then identified. ZTF J0923+4236 has a periodicity of the order of days, with more irregular transits. ZTF J0347-1802 appears to indicate a long duration transit of around 70 days. SDSS J0107+2107, ZTF J0328-1219 and SBS 1232+563 exhibit continuous transits on short time scales. Their transit depths vary, with SDSS J0107+2107 having much deeper transits exceeding even 25\%, ZTF J0328-1219 between 5-10 \% and the smallest dips are seen in SBSS 1232+563. A more detailed follow-up study for ZTF J0328-1219 revealed two periodicities, 9.937 and 11.2 hours, which were interpreted as debris clumps at two differing orbits.

Lastly, WD 1054-226 exhibits very regular transits every 25.02 hours. Within the fundamental period, there are dimming events which are separated by exactly 23.1 minutes, with depths of up to several \% and extraordinary night-to-night similarity.

The various associated values among these transit parameters are summarized in Table \ref{Tab:Transits} (cells are blank where data is not yet available). One can immediately notice an astounding variety of transit cases. It may eventually turn out that these represent the same general process, however observed at different evolutionary stages. It should also be noted that at least 3 of these transiting systems show long-term brightness variations on the scale of months to years \citep{AungwerojwitEtAl-2024}, presumably resulting from varying underlying processes which change the overall levels of dust production or dispersal.

\begin{table}[h!]
	\TBL{\caption{Transit characteristics around white dwarfs}\label{Tab:Transits}}
	{\begin{tabular*}{\textwidth}{@{\extracolsep{\fill}}@{}llll@{}}
			\toprule
			\multicolumn{1}{@{}l}{\TCH{system}} &
			\multicolumn{1}{l}{\TCH{period}} &
			\multicolumn{1}{l}{\TCH{duration}} &
			\multicolumn{1}{l}{\TCH{depth}}\\
			\colrule
			WD 1145+234 	& 4.5 hours & 5 minutes		& <50\%\\
			ZTF J0139+5245 	& 107 days	& $\sim$weeks 	& 20-45\%\\
			ZTF J0923+4236 	& 			& $\sim$days  	& \\
			ZTF J0347-1802 	& 			& 70 days		& \\
			SDSS J0107+2107 & 			& 				& <25\%\\
			ZTF J0328-1219 	&$\sim$10 hours&20-60 minutes&5-10\%\\
			SBSS 1232+563 	& 			& 				& <5\%\\
			WD 1054-226 	& 25 hours	& <23.1 minutes	& <5\%\\
			\botrule
	\end{tabular*}}{%
	}%
\end{table}

\subsection{X-ray}\label{SS:XRay}
The previously mentioned system G29-38 is the first and thus far only polluted WD system with X-ray detection \citep{CunninghamEtAl-2022}. From the measured X-ray luminosity, one may derive an instantaneous accretion rate onto the atmosphere, independent of stellar atmospheric models from which this value is typically inferred. 

\section{Theoretical interpretation of polluted white dwarf systems}\label{S:Theory}
Several theoretical models which were proposed in order to explain the observations of polluted WD disks are discussed below. After the discussion, a schematic depiction of these models is presented in Figure \ref{fig:Models}.

\subsection{Geometrically thin optically-thick disks}\label{SS:Classic}
The fiducial model of polluted WD disks by \cite{Jura-2003} considered debris disks that are analogous to the rings of Saturn in both scale and morphology. Such disks were imagined to be relatively quiescent and static, such that when subsequently observed, they were not supposed to manifest considerable changes on short human time scales. 

This model successfully reproduced the IR excess of most polluted WD, however as already indicated in Section \ref{SS:IRexcess}, there are at least three bright systems which do not adhere to this model. In recent years, it has become even clearer that such a simple view of WD disks could not possibly be correct. The IR variability discussed previously is incompatible with quiescent disks because there must be significant increase or decrease to the dust emitting area on relatively short time scales.

In the past, gas observations were most frequently attributed to sublimation of refractory materials along the inner part of the disk, subjected to the most intense stellar light. However, \cite{RocchettoEtAl-2015} showed that there is no increase in fractional luminosities of WD disks as they mature, despite the expected increase of the emitting area as the WD cools.

An additional problem is that gas cannot exist simultaneously in stable phase equilibrium with adjacent dust throughout the disk, which defies the observations. \cite{MetzgerEtAl-2012} point out that in an optically thick flat disk, the temperature of the solids is necessarily below the sublimation temperature for all but the innermost annulus of dust. If hotter gas exists at any time, atoms of gas should stick upon colliding with the surface of dust particles, resulting in condensation, the time scale of which is between a few to a few hundred orbital periods, implying full condensation in as little as a few hours.

The fiducial flat disk model also cannot be fully understood in the context of transits, which often point to material orbiting somewhat outside or even far outside the Roche radius. Hence, transiting systems must have, in addition to, or instead of a flat circular compact disk, also an eccentric body, bodies and/or dusty effluence to actually generate the dimming.

\subsection{Eccentric disks}\label{SS:Eccentric}
Since the original progenitors of debris disks start on highly eccentric orbits (Section \ref{S:Disruption}), a natural model to explain IR excess considers a strictly eccentric disk geometry. Although compatible with the IR flux, lack of ample long wavelength data does not enable to discern the exact eccentricity \citep{DennihyEtAl-2016}. \cite{NixonEtAl-2020} show how ring-like (Figure \ref{fig:Disruption}-a) dusty disks can account for the variability in the IR flux, as well as transits. This comes from the temporal spreading of localized clumps of dusty material, during the filling of the ring. Clumped dust would feature brightest in the IR near periastron, whereas evenly-spread dust rings result in a lower steady-state IR flux. A superposition of multiple rings is also possible, resulting in more complex IR patterns. 

\begin{figure}[h!]
	\centering
	\includegraphics[width=\textwidth]{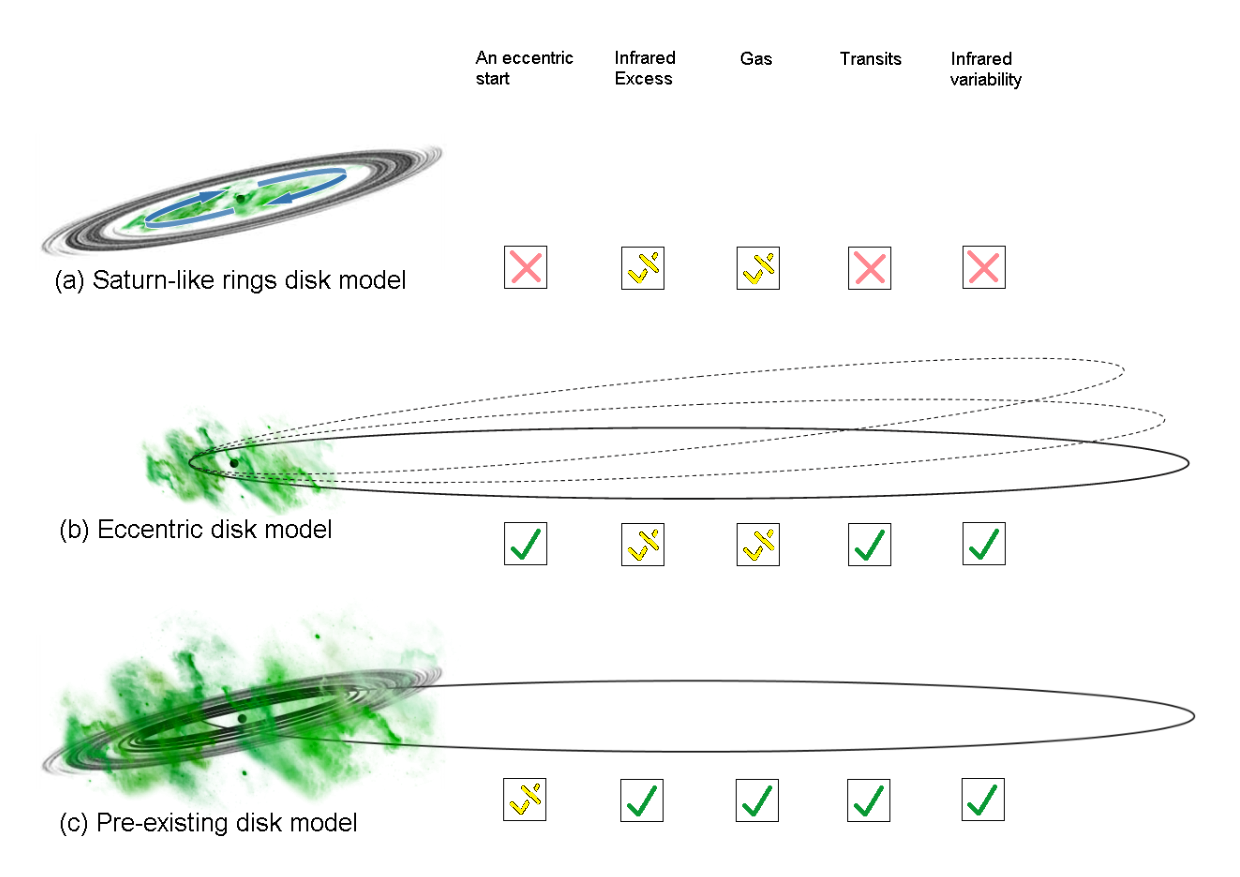}
	\caption{A schematic depiction (color coding - gray tones for debris consisting of fragments and green for gas) of different WD disk models, and their compliance with various constraints: (a) the fiducial WD disk model of Saturn-like rings is largely inapplicable; (b) the eccentric disk model probably applies to most polluted WDs, but is hard to reconcile with large accretion rates and no detected IR excess; (c) the pre-existing disk model probably complies with the largest number of constraints, however it is by design meant to address a limited fraction of polluted WDs.}
	\label{fig:Models}
\end{figure}

As previously shown in Section \ref{S:Disruption}, there are a few possible disk geometries following a tidal disruption, besides a ring. Additionally, the assumption of a dusty disk is incompatible with the initial size distribution of tidal fragments, which are approximately tens to hundreds of meters \citep{MalamudEtAl-2021}. \cite{BrouwersEtAl-2022} therefore consider eccentric disks that are not confined to a ring-like structure, and are also not strictly dusty. They are comprised instead of internally strong fragments, whose orbital energies are spread out according to the size and orbit of their progenitor asteroid. This results in an eccentric disk of interlaced elliptical annuli. Collisions between fragments occur when their relative orbits are altered sufficiently from their initial state, such that they begin to overlap. These changes are induced by the differential precession of the various annuli (see Figure \ref{fig:Models}-b where each annuli of tidal fragments with different separation is depicted by the dashed lines) as a result of general relativistic effects generated by the WD \citep{BrouwersEtAl-2022}, and potentially also from gravitational perturbations induced by a massive planet, exterior to the outer annuli of the debris disk \citep{VerasEtAl-2021}. Relative fragment-fragment collision velocities are then sufficiently large for catastrophic outcomes, grinding the fragments into progressively smaller sizes, and eventually dust. The collision rate can be directly translated into an accretion rate onto the star, by applying a radiation-related process called Poynting-Robertson (PR) drag, that causes dust particles to lose angular momentum and accrete. \cite{BrouwersEtAl-2022} show the modeled accretion rates to be similar to observation.

A potential problem is that such models predict large IR excess, however the opposite is observed - while there are many polluted WDs with relatively high inferred accretion rates, only a small fraction of order few \% has large detectable IR excess. \cite{BrouwersEtAl-2022} argue that this problem could be circumnavigated by more rapid dust accretion (than PR drag), perhaps as a result of gas drag or various magnetic phenomena that similarly extract angular momentum out of various sized dust particles or fragments. The problem with such propositions is that only about 1\% of polluted WDs are observed to have a gas component, and only about 10\% of WDs are highly magnetic.

Another channel for accretion requires a scenario-specific route involving the presence of a large planet \citep{VerasEtAl-2021,LiEtAl-2021,BrouwersEtAl-2022}. This planet first injects the progenitor asteroid to trigger a tidal disruption, and the ensuing disk of debris must cross the planet's orbit again, at least partly, and fragments can be scattered by gravitationally interacting with the planet. \cite{LiEtAl-2021} show that only large asteroids of at least 100 km in size contain enough mass in order to grind them down very quickly, after which PR drag can effectively circularize the debris and form a more circular and compact debris disk around the WD. Otherwise, circularization and accretion are slow to occur.

\subsection{Pre-existing compact disks}\label{SS:Compact}
The pre-existing compact disk model \citep{MalamudEtAl-2021} is an extension of the fiducial flat disk model, inspired by the role of collisions which were suggested to lead to observed variable IR flux \citep{SwanEtAl-2020}. This model suggests that many WDs with observed IR excess are those which have a pre-existing, nearly circular compact disk within the WD's Roche radius, in compliance with having a baseline IR excess. In order to explain the outlier IR excess cases, the IR variability, the eccentric transits and also ongoing gas production, highly eccentric streams of tidal fragments are first produced via occasional tidal disruptions, and are therefore intersecting the existing compact disk. Extreme hyper-velocity collisions lead to fragment erosion, scatter of dusty material, and gas generation.

These collisions also act to circularize the tidal fragments, thus bridging the gap between the eccentric disk and flat disk models, previously outlined. The characteristic time scale to circularize fragments is much smaller for small tidal fragments, so they are the first to orbitally shrink and conjoin the compact disk. Given a likely size-distribution of the tidal stream fragments, gas production is first triggered by continuous collisions of small fragments, yielding an observable signature of ongoing early circularization. But as the smaller fragments circularize, the gas production originates from increasingly larger fragments and thus it becomes increasingly intermittent and harder to observe. Gas production finally halts completely when either all the tidal fragments are circularized or the compact disk has dissipated (which depends on the ratio between the tidal stream and compact disk masses). Thus, the model offers a natural explanation for the smaller fraction of gaseous material in polluted WDs with IR excess.

IR variability and the outlier IR excess cases may be explained by a second, optically thin dusty halo component. The explanation of the transit observations is similar to that invoked by the eccentric disk model. Both the pre-existing compact disk and the eccentric disk models are quite recently added to the literature, and include various aspects that require future investigation. This makes the theoretical study of polluted WDs one of the most cutting-edge and fastest developing areas in astrophysics.

\section{Summary}\label{S:Summary}
The study of polluted white dwarf debris disks involves the analysis of material accreting onto white dwarfs from surrounding planetary bodies. Over the past two decades, advancements in observational techniques, in space-based telescopes and ground-based facilities, have led to ongoing discovery of polluted systems in ever growing numbers. Their disks too, while still relatively rare, are also continually found and characterized.

Theoretical models have been developed in order to explain the presence of debris disks and the processes leading to the pollution of white dwarfs. These models take into account various interactions with planetary bodies.

Despite these successes, the field has faced challenges in understanding the exact mechanisms responsible for the observed pollution, the diversity of compositions found in the accreted material and also the astonishing variety of different systems with debris disks, many exhibiting dissimilarities in their transits, gas or infrared excess signals.

There are still plenty of open questions. However, the future of the field looks extremely promising. Large scale surveys will continue to discover new polluted white dwarfs. These surveys will provide a broader understanding of the prevalence and characteristics of such systems. Advanced telescopes like the James Webb Space Telescope and others, will significantly enhance our ability to study specific polluted white dwarfs and probe their disks in an increasing level of detail. Several missions are planned in order to find and characterize exoplanets around a large number of stars, potentially including white dwarfs as well. Coordinated observations across different wavelengths will provide a more comprehensive view of particularly interesting systems. Continued theoretical developments will refine our understanding of the processes governing the formation and evolution of polluted white dwarf debris disks. This includes improved models for interactions between different types of planetary bodies and various types of disks, the role of collisions and of tidal disruptions.

In summary, the field of polluted white dwarf debris disks has seen notable progress in the past two decades, with advancements in both observations and theoretical modeling. The future holds similar promise, with potential breakthroughs from new instruments and missions that will enhance our ability to study these intriguing astronomical phenomena.

%\begin{quote}
%\quotehead{Quotehead}
%The foundation for any model is a set of conservation principles. For mesoscale atmospheric models, these principles are conservation of mass, conservation of heat, conservation of motion, conservation of water, the conservation of other gaseous and aerosol materials, and an equation of state.
%\source{--source}
%\end{quote}

%\begin{ack}[Acknowledgments]
%...
%\end{ack}

%\seealso{article title article title}
\newpage
\bibliographystyle{Harvard}
\bibliography{reference}

\end{document}